\documentclass[epj]{svjour}

\usepackage{graphics}

\begin{document}

\title{Ground-state Properties of a Supersolid in RPA}
\subtitle{}
\author{A. J. Stoffel\inst{1,2} \and M. Gul\'{a}csi\inst{1,2}}   
\institute{
Max-Planck-Institute for the Physics of Complex Systems, 
D-01187 Dresden, Germany \and
Nonlinear Physics Centre, Australian National University,
Canberra, ACT 0200, Australia
}

\date{Received: date / Revised version: date}

\abstract{
We investigate the newly discovered supersolid phase by solving in random-phase approximation
the anisotropic Heisenberg model of the hard-core boson ${}^4$He lattice at zero temperature.
We include nearest and next-nearest neighbor interactions and calculate exactly all pair correlation
functions in a cumulant decoupling scheme. We demonstrate the importance of vacancies and 
interstitials in the formation of the supersolid phase. The supersolid phase is characterized by
strong quantum fluctuations which are taken into account rigorously. Furthermore we confirm 
that the superfluid to supersolid transition is triggered by a collapsing roton minimum however 
is stable against spontaneously induced superflow, i.e. vortex creation.
}
\maketitle
\section{Introduction}
The counterintuitive idea of a superflow in a solid, later coined
supersolidity was firstly conjectured in 1969 by Andreev\cite{andreev} 
and in 1970 seized by Leggett and Chester\cite{chester,leggett}. 
From a theoretical 
point of view, supersolidity is a state of matter characterized by 
simultaneous off-diagonal (ODLRO) and diagonal long range order (DLRO). 
It was speculated that such a  phase exists because
vacancies and other defectons are non-localized and will
Bose condense at sufficiently low temperature.  
Still most physicists remained critical of the notion 
as several experiments failed to produce any evidence of this state.
Finally in 2004 Kim and Chan \cite{kim1,kim2}
measured a tiny superflow in solid helium 
at temperatures below T=0.2 K, expressed by non-classical rotational inertia
in a torsional oscillation experiment, and
thus proved the existence of the supersolid state. 
This landmark experiment rekindled vast interest in the 
supersolid state and   
subsequently many new theories  and numerical quantum 
Monte-Carlo calculations supporting the existence of supersolidity were proposed. 
However the true nature of the supersolid phase still remains
obscure. Numerous follow-up experiments  managed to shed 
light on the matter but the relevance of ${}^3$He impurities and 
especially the nature of the unconventional  
normal solid to supersolid transition resembling  the 2D Kosterlitz-Thouless transition 
is still being debated.\newline
Recent experiments\cite{aoki} raised new questions as it was found that the
supersolid phase exhibits a hysteresis, where the superfluid signal depends
on the chronology of variation of temperature and in the 
amplitude of the rotational oscillation. An other recent experiment\cite{day} 
detected a change in the elastic properties of solid helium. 
The change of the elastic moduli bears a 
remarkable resemblance with the supersolid signal.\newline
However,  despite sophisticated numerical methods and advanced 
theories such as vortex liquids\cite{anderson} and 
superglass states\cite{superglass} 
we believe that there remains a gap in the range of theories 
of the supersolid phase. In this paper we intend to fill this gap and
present a theory of supersolidity in a quantum lattice gas (QLG) model beyond classical mean-field. 
We follow the approach of K.-S. Liu and M.E. Fisher\cite{fisher1}
and map the QLG model to
the anisotropic Heisenberg model. 
The method of Green's functions proved to be very successful 
in the description of ferromagnetic and anti-ferromagnetic states and 
we use this method to investigate the supersolid phase which 
corresponds to a canted anti-ferromagnetic phase.

The emergent third order Green's functions in the random-phase approximation (RPA)  
are broken down using the cumulant decoupling to yield 
 a closed set of equations.
Quantum fluctuations at zero temperature result in vacancies
and interstitials present even at zero temperature and in the supersolid
phase the net vacancy density is therefore non zero. The supersolid 
phase is characterized by  Bose condensation of the vacancies as well
as the interstitials and thus both will contribute to superfluidity. 
Interestingly, the major contribution comes from vacancies.
Also, our model confirms propositions that the superfluid to supersolid transition
is triggered by a collapsing roton minimum\cite{rotonmin,zhao}.
 Nonetheless our solution shows,
 contrary to earlier results
that this transition is stable against spontaneously induced superflow.
\newline
The paper is organized as follows: In Section \ref{sec:GenHam}
and \ref{sec:aHM} we
introduce the generic Hamiltonian of a bosonic many 
body system and discretize it to a  
quantum lattice gas model. This model is equivalent 
to the anisotropic Heisenberg model 
in an external field and we will identify the corresponding 
phases. 
In the following two sections we derive basic thermodynamic 
properties relevant at zero temperature 
and discuss the excitation spectrum of the spin waves in 
the superfluid and the supersolid phases.
The relevance of quantum fluctuations is discussed in Section 
\ref{sec:qfluc}, where we provide justification of results
briefly reported elsewhere \cite{EPL}.
Finally in the last two sections we discuss key properties 
of the supersolid phase and their occurrence within the 
example of three different sets of coupling constants.
\newline
\section{Generic Hamiltonian}\label{sec:GenHam}
Apart from possible ${}^3$He impurities the supersolid Helium-4  state is 
a bosonic system and the generic Hamiltonian for such systems 
in the language of second quantization is given by:
\begin{eqnarray} \label{genbosH}
H&=&\int d^3 x \psi^{\dagger}(x)( -\frac{1}{2 m }\nabla^2+
\mu) \psi(x)  \nonumber\\
 &&+\frac{1}{2}\int d^3x d^3x'\psi^{\dagger}(x)\psi^{\dagger}(x') V(x-x')
 \psi(x)\psi(x')\nonumber\\
\end{eqnarray}
where $\psi^{\dagger}(x)$, the particle creation operator and 
 $\psi(x)$ , the corresponding destruction operator obey the usual bosonic 
commutator relations.
Hamiltonians in three dimensions such as in Eq. (\ref{genbosH}) are not solvable 
even for elementary potentials $V(x)$ such as the Dirac delta distribution.
Therefore we are induced to introduce further approximations.
An approximation which proved particularly successful 
for the description of liquid Helium is know
as the Quantum Lattice Gas model and was first introduced
by Matsubara and Matsuda\cite{matsubara}.
\newline
In the quantum lattice gas model one works with a space lattice of discrete lattice points
rather than the continuum.  This approximation shows 
to be very useful for the supersolid state
as the spatial discretization of the this model
serves as a natural frame for the crystal lattice
of (super)-solid helium. This procedure significantly
simplifies the problem of breaking 
translational invariance symmetry for states
that exhibit diagonal long range order. 
In this way this model gives the easiest possible access
 to analyze states that exhibit both diagonal 
and off diagonal long range order simultaneously. 
Also in this model no specific knowledge
of the density distribution of the atoms is needed.
\newline
According to Matsubara and Tsuneto\cite{matsuda} the generic 
Hamiltonian Eq. (\ref{genbosH}) in 
the discrete lattice model reads:
\begin{eqnarray}\label{hcbhubb}
H=\mu \sum_i n_i+
\sum_{ij}u_{ij}(a_i^{\dagger}-a_j^{\dagger})
(a_i-a_j)
+\sum_{ij} V_{ij} n_i n_j\nonumber\\
\end{eqnarray}
Here $u_{ij}$ are non-zero for nearest neighbor and next nearest neighbor hopping
and otherwise zero. The values of $u_{nn}$ and $u_{nnn}$ are  such
that  the kinetic energy is isotropic up to the 4th order.
In the case of a BCC lattice (two interpenetrating SC lattices) the matrix
elements are given by:
\begin{eqnarray}
u_{nn}=\frac{2}{3}\frac{1}{4 m a^2}\nonumber\\
u_{nnn}=\frac{1}{3}\frac{1}{4 m  a^2}
\end{eqnarray}
As atoms do not penetrate each other there 
can exist only one atom at a time on a lattice site 
and consequently
$a^{\dagger}$ and $a$ are the creation and annihilation 
operators of a hard core boson commuting on different lattice sites:
\begin{eqnarray}
[a^{\dagger}_i,a^{\dagger}_j]_{-}=
[a_i,a_j]_{-}=
[a_i,a^{\dagger}_j]_{-}=0 \; (i\neq j)
\end{eqnarray}
but obey the anti-commutator relations on identical sites:
\begin{eqnarray}
[a^{\dagger}_i,a^{\dagger}_i]_{+}=
[a_i,a_i]_{+}=0   \nonumber \\
 \left[a_i,a^{\dagger}_i\right]_{+}=1 
\end{eqnarray}
Equation (\ref{hcbhubb}) is the Hubbard model in 3 dimensions 
for hard core bosons. Due to the unusual statistics
of hard core bosons, there does not exist
a Wick's theorem for their operators and 
the common formalism of pertubative field theory is not applicable.
Hence in the following chapter we transform the model to an 
equivalent spin model namely the anisotropic Heisenberg model.
\section{Anisotropic Heisenberg Model}\label{sec:aHM}
\begin{figure}
\centering
\resizebox{5 cm}{!}{
\includegraphics*{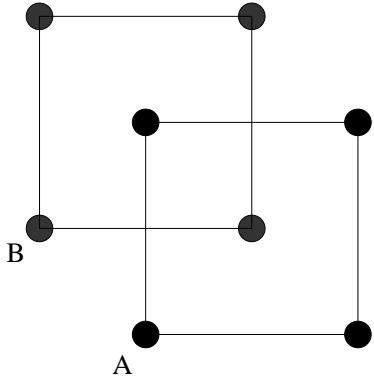} }
\caption{\label{fig:one} The BCC lattice consists of two interpenetrating 
SC sub-lattices. In the perfect solid phase one sub-lattice (i.e. sub-lattice A) serves 
as a lattice of on-site centers and is occupied while sub-lattice B represents the empty
interstitial and is vacant.}
\end{figure}

It is well known that the operators of 
hard-core bosons obey the same SU(2) 
algebra as spin $S=1/2$ particles do. 
Therefore it is feasible to replace the 
creation and annihilation operators
by spin operators.
\begin{eqnarray}
a^{\dagger}_j=S^x_j-i S^y_j \nonumber  \\
a_j=S^x_j+i S^y_j    \nonumber \\
n_j=\frac{1}{2}- S^z_j.
\end{eqnarray}
It is easily apprehensible that the usual lie algebra for
spin 1/2 particles preserves the mixed commutation/anti-commutation relations
for hard-core bosons. 
This substitution transforms the hard-core bosonic Hubbard 
model into a spin model:
\begin{eqnarray}
\lefteqn{H=\mu\sum_i (\frac{1}{2}-S^z_i)}\nonumber\\
&&+\sum_{ij}u_{ij}
(1-S^z_i-S^z_j-2S^x_i S^x_j-2S^y_iS^y_j) \nonumber \\
&&+\sum_{ij}V_{ij}(\frac{1}{4}-\frac{S^z_i}{2}-\frac{S^z_j}{2}+S^z_iS^z_j)
\end{eqnarray}
if we adjust the notation to conform with the usual standards
of spin models, it becomes evident that the resulting 
Hamiltonian is the anisotropic Heisenberg model :
\begin{eqnarray}
\lefteqn{H=-h^z \sum_i S^z_i}\nonumber\\
&&-\sum_{ij}J^{\|}_{ij}S^z_iS^z_j-
\sum_{ij}J^{\top}_{ij}(S^x_i S^x_j+S^y_iS^y_j) 
\end{eqnarray}
with:
\begin{eqnarray}
J^{\|}_{ij}=-V_{ij}\nonumber \\
J^{\top}_{ij}=2 u_{ij}
\nonumber\\
h^z=\mu-\sum_{j}J^{\top}_{ij}
+\sum_{j}J^{\|}_{ij}
\end{eqnarray}
The Hubbard model as well as the anisotropic Heisenberg
model are defined on a discrete lattice and one may ask to what extend
a specific choice of lattice geometry will
 affect the physical properties of the system.
While the quantitative results
 certainly depend on the lattice geometry
we can safely assume that qualitative properties,
 such as phase transitions and critical
constants will not change for different lattices
 as long as no frustration effects are evoked. 
 Therefore we may safely choose, in order to avoid
 unnecessary complications, 
a simple lattice geometry and an obvious
 choice are two interpenetrating simple cubic sub-lattices
 which together form a 
BCC lattice, see Figure \ref{fig:one}. 
Defining two sub-lattices gives us the possibility to establish the DLRO
of solids in a natural way: sub-lattice A represents the centers of the 
${}^4$He ions, hence it coincides with the ion lattice. Sub-lattice B
defines the interstitial, the space in-between those atomic centers.
In the liquid phases of course the occupation number on both 
sub-lattices is equal as there is no spatial density variation.
Table \ref{tab:one} charts the various magnetic phases and 
identifies the corresponding phases of the ${}^4$He system. 
According to the spin configurations we call the four magnetic phases 
ferromagnetic, canted ferromagnetic, canted anti-ferromagnetic and 
anti-ferromagnetic phases which we abbreviate by FE, CFE, CAF and AF. 
The order parameters, $m_1$
for off-diagonal long range order and  $m_2$
for diagonal long range order
in the magnetic system are defined by:
\begin{eqnarray}
m_1=\langle S^x_A\rangle+\langle S^x_B\rangle \nonumber \\
m_2=\langle S^z_A\rangle-\langle S^z_B\rangle
\end{eqnarray}
In the following we will use these order parameters to identify 
the phases within the classical mean-field approximation as was
derived by Matsuda and Tsuneto \cite{matsuda} and extended to finite temperature by 
Fisher and Liu \cite{fisher1} as well as in the novel random-phase approximation.
\newline\newline
\begin{table}
\caption{ Possible magnetic phases and the corresponding 
phases of the Hubbard model. All Phases are defined by their
long range order. The columns, from left to right, are the spin
configurations, magnetic phases, ODLRO, DLRO and corresponding 
${}^4$He phases, respectively.\label{tab:one}}
\center
\begin{tabular}{|c|c|c|c|c|}
\hline
&&&&\\
$\uparrow\uparrow$ & FE & No & No & Normal Liquid \\
&&&&\\
$\nearrow\nearrow$ & CFE & Yes & No & Superfluid \\
&&&&\\
$\nearrow\swarrow$ & CAF  & Yes & Yes & Supersolid \\
&&&&\\
$\uparrow\downarrow$& AF & No & Yes & Normal Solid \\
&&&&\\
\hline
\end{tabular}
\end{table}
\section{Green's Function}\label{sec:GF}
The anisotropic Heisenberg model in an external field is, 
not least due  to absence of O(3) symmetry difficult to solve. However,
in the context of supersolidity, investigating the CAF  phase, it has been 
solved in a classical mean-field approximation\cite{fisher1,matsuda}.
The anisotropic Heisenberg Hamiltonian in the classical
mean-field approximation is obtained by substituting the 
spin-1/2 operators with their respective 
expectation values:
\begin{eqnarray}
 \lefteqn{ H_{MF}=- h^z  (\langle S^z_A \rangle +
    \langle S^z_B \rangle)}\nonumber\\
&&-2 J^{\|}_1 \langle S^z_A\rangle\langle S^z_B\rangle
-J^{\|}_{2}(\langle S^z_A\rangle \langle S^z_A \rangle + 
\langle S^z_B\rangle \langle S^z_B\rangle) \nonumber\\
&&-2 J^{\top}_1 \langle S^x_A\rangle\langle S^x_B\rangle
-J^{\top}_{2}(\langle S^x_A\rangle \langle S^x_A \rangle 
+ \langle S^x_B\rangle S^x_B\rangle ), 
\label{HMF}
\end{eqnarray}
where $J^{\|}_1=q_1 J^{\|}_{i\in A j\in B} $, $J^{\|}_2=q_2 J^{\|}_{i\in A j\in A}$
, $J^{\top}_1=q_1 J^{\top}_{i\in A j\in B}$ and $J^{\top}_2=q_1 J^{\top}_{i\in A j\in A}$,
$q_1=6$ and $q_2=8$ are the number of nearest and next nearest neighbors.
\begin{figure}
\centering
\resizebox{8.5 cm}{!}{
\includegraphics*{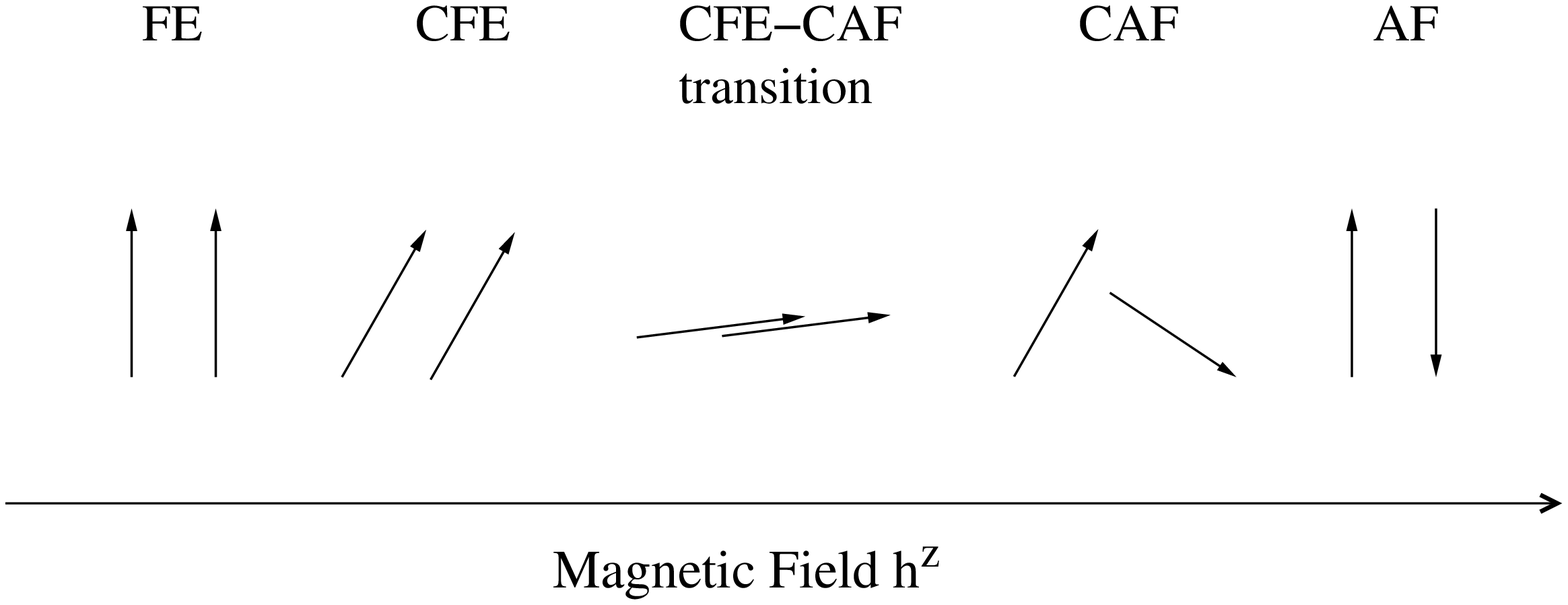} }
\caption{\label{fig:spinconf}The possible mean-field phases of the 
anisotropic Heisenberg model on a bipartite lattice
with external field $h^z$.}
\end{figure}
Although the classical mean-field theory is quite insightful and gives
an accurate description of the variously ordered phases its fails to take
quantum fluctuations and quasi-particle excitations into account.
Hence, in order to overcome these shortcomings we derive a fully
quantum mechanical approximation and solve the anisotropic Heisenberg
model in the random-phase approximation which is based on the 
Green's function technique.
At zero temperature the retarded and advanced Tyablikov \cite{bogo,tyablikov} commutator Green's 
function  defined in real time are:
\begin{eqnarray}\label{GFdef}
G^{\mu\nu}_{ij_{Ret}}(t)=-i\theta(t)\langle n_0|[S^{\mu}_i(t),S^{\nu}_j] |n_0 \rangle \nonumber \\
G^{\mu\nu}_{ij_{Adv}}(t)=i\theta(-t)\langle n_0|[S^{\mu}_i(t),S^{\nu}_j] |n_0 \rangle
\end{eqnarray}
Here $|n_0 \rangle$ is the normalized Heisenberg ground state,
$\mu$ and $\nu$ are elements of $\{x,y,z\}$ and $i$ and $j$
denote the lattices sites.
The most successful technique of solving many body Green's function involves
the method of equation of motion which is given by:
\begin{eqnarray}\label{eq_of_m}
i \partial_t G^{xy}_{ij_{Ret}}(t)= \delta(t)\langle[S^x_i,S^y_j]\rangle
-i\theta(t)\langle[[S^x_i,H],S^y_j]\rangle \nonumber \\
i \partial_t G^{xy}_{ij_{Adv}}(t)= \delta(t)\langle[S^x_i,S^y_j]\rangle
+i\theta(-t)\langle[[S^x_i,H],S^y_j]\rangle \nonumber\\
\end{eqnarray}
The commutator $[S^x_i,H]$ can be eliminated by using the 
Heisenberg equation of motion giving rise to 
higher, third order Green's functions on the RHS. 
In order to obtain a closed set of equations we 
apply the cumulant decoupling procedure and as a consequence 
the third order Green's functions split into 
product terms of single operator expectation values
and second order Green's functions.
The cumulant decoupling \cite{brown} is based on the assumption that
the last term of the following
equality is negligible:
\begin{eqnarray}
\lefteqn{\langle \hat{A}\hat{B}\hat{C} \rangle =} \nonumber\\
&&\langle \hat{A}\rangle\langle \hat{B}\hat{C} \rangle +
\langle \hat{B}\rangle\langle \hat{A}\hat{C} \rangle \nonumber\\ 
&&+\langle \hat{C}\rangle\langle \hat{A}\hat{B} \rangle 
-2\langle \hat{A}\rangle\langle \hat{B}\rangle\langle\hat{C} \rangle
\nonumber \\
&&+\langle (\hat{A}-\langle \hat{A} \rangle) 
(\hat{B}-\langle \hat{B} \rangle)
(\hat{C}-\langle \hat{C} \rangle) \rangle
\end{eqnarray}
This decoupling scheme leads to a closed set of six equations that
determine the six Green's 
functions, corresponding to three spin components on two sub-lattices.
\newline
The cumulant decoupling scheme introduces mean-fields of the spins operators
which have to be calculated in a self-consistent manner.  
The Green's functions gives us the possibility to calculate 
a set of two self-consistency equations. However, in the supersolid and superfluid case
the off-diagonal long range order; i.e. non-zero 
transversal fields $\langle S^x_A\rangle$ and
$\langle S^x_A\rangle$
increase the degrees of freedom in number by two and therefore 
analytical properties of the commutator Green's functions pose two additional 
constraints on the spin-fields given by:
\begin{eqnarray} \label{mf2}
   h^z+2 \langle S^z_A \rangle (J_2^{\|}-J_2^{\top})+2 \langle S^z_B \rangle J_1^{\|}=
   2 J_1^{\top}\frac{\langle S^x_B\rangle }{\langle S^x_A\rangle }\langle S^z_A\rangle 
   \nonumber \\
   h^z+2 \langle S^z_B\rangle  (J_2^{\|}-J_2^{\top})+2 \langle S^z_A \rangle J_1^{\|}=
   2 J_1^{\top}\frac{\langle S^x_A\rangle }{\langle S^x_B\rangle }\langle S^z_B \rangle 
\end{eqnarray}
These equations are quite important and
also hold in the classical mean-field approximation, 
where the state of the system is 
defined by these equations together with\newline 
 $\sqrt{\langle  S^x_A\rangle^2 +\langle S^z_A \rangle^2 }=\frac{1}{2}$
and $\sqrt{\langle  S^x_B\rangle^2 +\langle S^z_B \rangle^2 }=\frac{1}{2}$.
Equation (\ref{mf2}) also determines possible 
second order phase transitions as was shown my Matsuda and Tsuneto\cite{matsuda}.
The normal fluid to superfluid (FE-CFE) second order transition is located at:
\begin{equation}\label{tpfsf}
h^z_{FE-CFE}=J^{\top}_1+J^{\top}_2-J^{\|}_1-J^{\|}_2
\end{equation}
For the normal solid to supersolid (CAF-AF) the classical mean-field approximation
 yields:
\begin{equation}\label{tpiaf}
h^z_{CAF-AF}=\sqrt{(-J^{\|}_1+J^{\|}_2-J^{\top}_2)^2-(J^{\top}_1)^2}
\end{equation}
and the critical magnetic field $h^z$ (corresponds to the chemical potential $\mu$ in
the language of the QLG) for the superfluid to supersolid (CFE-CAF) transition is:
\begin{eqnarray}\label{tpisf}
h^z_{CFE-CAF}=\frac{J^{\|}_1+J^{\|}_2-J^{\top}_1-J^{\top}_2}
{J^{\|}_1-J^{\|}_2-J^{\top}_1+J^{\top}_2}\times\nonumber \\
\sqrt{(-J^{\|}_1+J^{\|}_2-J^{\top}_2)^2-(J^{\top}_1)^2}
\end{eqnarray}
For a particular choice of coupling constants all four phases 
will exists when, see Figure \ref{fig:spinconf}:
\begin{eqnarray}
h^z_{FE-CFE}>h^z_{CFE-CAF}>h^z_{CAF-AF}
\end{eqnarray}
The three equations for the critical fields, Eq. (\ref{tpfsf}), Eq. (\ref{tpiaf})
 and Eq. (\ref{tpisf}),
are derived for in
classical mean-field approximation. However as Equation (\ref{mf2}) also holds in the 
random-phase approximation these values give a good indication where the 
actual transitions take place. Nevertheless, due to depletion of the spin-fields 
caused by quantum fluctuations the actual values are slightly lower, 
see Section \ref{sec:disc}.
\section{Thermodynamic Properties}\label{sec:thermo}
In the first section we have seen that the grand-canonical 
QLG Hamiltonian, where the number of particles are variable
corresponds to the canonical anisotropic 
Heisenberg Hamiltonian with fixed number of spins. 
Therefore there exists following relation between any thermodynamic
potential defined in the QLG model and the anisotropic Heisenberg model:
\begin{equation}
\Theta_{QLG}-\mu N=\Theta_{Heisenberg}
\end{equation}
Here $\Theta$ refers to an arbitrary thermodynamic potential.  \newline
The self-consistency equations, derived in the previous sections, 
determine the spin fields of the anisotropic Heisenberg model in the various phases, but 
in regions of $h^z$ where more than one solution may exist, we have
to compare internal energies to select the 
true ground state. Intuitively, we might want
 to compute the internal energy by calculating 
 the expectation value of the Hamiltonian as:
\begin{eqnarray}
H=h^z  \sum_i \langle S^z_i \rangle +\sum_{ij}J^{\|}_{ij}\langle S^z_iS^z_j \rangle 
\nonumber \\
+
\sum_{ij}J^{\top}_{ij}(\langle S^x_i S^x_j \rangle +\langle S^y_iS^y_j \rangle )
\end{eqnarray}
where the correlation functions can be derived from the corresponding Green's functions, Eq. 
(\ref{GFdef}).
Unfortunately this approach will yield incorrect and inconsistent results as the Green's function
derived with the
cumulant decoupling is not an exact solution of the anisotropic Heisenberg 
Hamiltonian but rather the solutions of an unknown, effective model, which is 
an approximation
of the Heisenberg model. As we do not know the exact 
form of this effective model we
have to resort to fundamental thermodynamic relations to integrate the energy.
At T=0 following equality holds:
\begin{equation}\label{freng}
\frac{\partial U}{\partial h^z}=\frac{\langle S^z_A\rangle 
+\langle S^z_B\rangle }{2}
\end{equation}
This equation allows us to determine if the critical fields given by 
Eq. (\ref{tpfsf}), Eq. (\ref{tpiaf}) and  Eq. (\ref{tpisf}) really refer
 to second order phase transitions.
As the z-component of the spin is decreased in the canted ferromagnetic phase due to
the onset of the transversal field this phase is energetically favorable over  
the ferromagnetic phase. Similarly the canted anti-ferromagnetic phase
has lower energy than the anti-ferromagnetic phase as 
the total magnetization in z-direction in the canted anti-ferromagnetic
phase is somewhat higher due
to the increasing influence of the ferromagnetic term $h^z  \sum_i S^z_i$.
Hence, the phases as depicted in Figure \ref{fig:spinconf} are real.
\newline
Unfortunately Eq. (\ref{freng}) only allows one to determine the energetically
favorable state when the possible transition point is known,
such as in second order phase transitions. But we can not use the 
relation to determine possible first order transitions 
and unfortunately this shortcoming is only resolvable 
by extending the formalism to  finite T.
\newline
Thermodynamic relations connect various macroscopic quantities 
and we will use them to obtain observable properties. 
Although the external magnetic field $h^z$ in the spin model
is an observable the corresponding  quantity in the QLG model,
 namely the chemical potential is not. Therefore we are interested to 
attain a formula for the  pressure associated with a certain chemical 
potential. 
The relationships is most easily derived from the following Maxwell relation:
\begin{equation}\label{pressa}
\left( \frac{\partial P}{\partial \mu}\right )_{T,V} =\left( \frac{\partial N}
{\partial V}\right )_{T,\mu}=\frac{\#_{lattice\_sites}}{V}(1-\epsilon)
\end{equation}
where $\epsilon:=\langle S^z_A\rangle +\langle S^z_B\rangle$.
These are the basic thermodynamic properties that 
we will use in the further discussion and in principle
all other properties can be derived from the internal energy U.
\section{Excitation spectrum}\label{sec:exc}
\begin{figure}
\centering
\resizebox{8.5 cm}{!}{
\includegraphics*{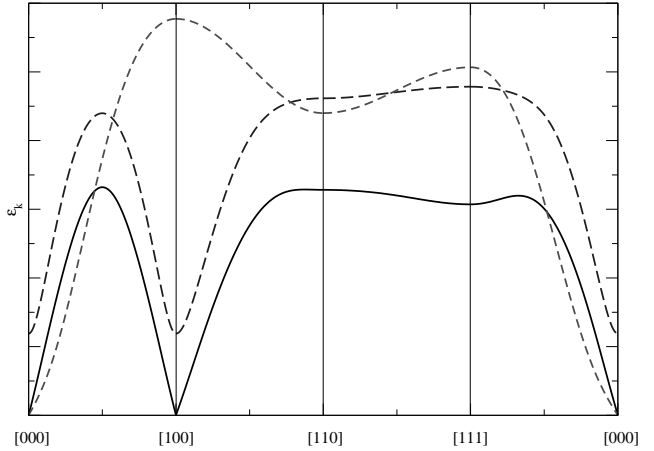} }
\caption{\label{fig:ssdr} Excitation spectrum of the anisotropic 
Heisenberg model with $J^{\top}_1=1.498 K$, $J^{\top}_2=0.562 K$,
$J^{\|}_1=-3.899 K$ and $J^{\|}_2=-1.782 K$ for the 
supersolid phase (solid line and dashed line) at $h^z=0.65$ and the
superfluid phase (long dashed line) at $h^z=7.46$. 
Here $[000], [100], [110], [110]$ refer to the various  points
of the first Brillouin zone.} 
\end{figure}

The excitation spectrum in the ferromagnetic phase as well as in the 
anti-ferromagnetic phase feature the well known,
gaped magnon excitation with
quadratic $k$ dependence in the $k\rightarrow 0$ limit.
In the canted ferromagnetic (superfluid) phase the spin-wave 
excitation spectrum is comprised, due to spontaneously
broken U(1) symmetry, of the gapless linear 
Goldstone phonons.\newline 
Additionally the canted anti-ferromagnetic (supersolid)
phase exhibits a second branch which accounts for 
the broken translational symmetry.
This second branch goes quadratic with $k$ 
in the long wavelength limit and
has a gap:
\begin{eqnarray}
\Delta=
[J^{\top}_1 2(J^{\|}_1-J^{\|}_2+J^{\top}_2)
\langle S^x_A \rangle^2 \langle S^x_B\rangle^2\nonumber \\
+ {J^{\top}_1}^2(\frac{\langle S^x_A \rangle^2+\langle S^z_A \rangle^2}
   {\langle S^x_A \rangle}\langle S^x_B \rangle^3+
\frac{\langle S^x_B \rangle^2+\langle S^z_B \rangle^2}
   {\langle S^x_B \rangle}\langle S^x_A \rangle^3\nonumber \\
+2\langle S^x_A \rangle\langle S^x_B \rangle
\langle S^z_A \rangle\langle S^z_B \rangle
) 
]^{\frac{1}{2}}\quad\quad
\end{eqnarray}
In Figure \ref{fig:ssdr} the quasi-particle excitation spectrum is 
plotted for the superfluid (CFE) and the supersolid (CAF) 
phases. In the supersolid (CAF) phase the 
excitation energy 
reaches zero at the edge (point [100]) of the first Brillouin zone.
The corresponding spin-waves refer to oscillations with
a $\pi$-phase shift between different sub-lattices.
Hence, on a single sub-lattice the spin-wave looks
like a zero wave-number mode.
It was recently suggested \cite{rotonmin} that the superfluid to 
supersolid transition is triggered by a collapsing roton minimum.
This assumption is supported by the present model; here the 
 transition to the supersolid phase
takes place when the excitation spectrum 
goes soft at [100].  
The dispersion relation in the superfluid (CFE) phase is given by:
\begin{eqnarray}\label{sfdr}
 \omega(k)=2 \{(J^{\top}_1 (\gamma_1(k)-1)+J^{\top}_2 (\gamma_2(k)-1))\times\nonumber\\
  \left[ \langle S_z\rangle^2 (J^{\top}_1
   (\gamma_1(k)-1)+J^{\top}_2 (\gamma_2(k)-1))-\right. \nonumber \\
   \left. \langle S_x\rangle^2 (J^{\top}_1+J^{\top}_2-J^{\|}_1
   \gamma_1(k)-J^{\|}_2 \gamma_2(k)) \right] \}^{1/2},
\end{eqnarray}
where $\gamma_1(k)$ and $\gamma_2(k)$ denote the standard
lattice generating functions of a BCC lattice.
From this equation we can see that the energy possibly goes to 
zero at [100] (corresponds to $\gamma_1(k)=-1$ and $\gamma_2(k)=1$)
when the following condition is fulfilled:
\begin{equation}{\label{iscr1}}
J^{\top}_1+J^{\top}_2+J^{\|}_1-J^{\|}_2<0
\end{equation}
Hence we obtained a further condition (supplementary to Eq. (\ref{tpisf}))
for the existence of the superfluid to supersolid transition.
\newline
Equation (\ref{sfdr}) allows for the existence of a second region of the 
reciprocal space where the 
dispersion relation might  go soft. 
For $\gamma_1(k)=0$ and $\gamma_2(k)=-1$ which corresponds to
[111] 
we obtain following condition:
\begin{equation}\label{iscr2}
\frac{-2 J^{\top}_2}{
J^{\top}_1+J^{\top}_2+J^{\|}_2}>0
\end{equation}
It was also conjectured that the superfluid phase
in the vicinity of the superfluid to supersolid
transition is unstable against spontaneously induced 
superflow and superflow associated with vortices.
Therefore we investigate how the present model
reacts to induced superflow. A net superflow
is either given by a moving condensate which results in  
a gradient of the phase
of the wave-function, or equivalently by 
a moving environment while the condensate stays at rest.
The latter is obtained by an additional term in the Hamiltonian:
\begin{eqnarray}
H_1=\int d^3 x \psi^{\dagger}(x)(i\mathbf{v_n}\cdot \nabla
) \psi(x)  
\end{eqnarray}
which corresponds to following term in the anisotropic Heisenberg model:
\begin{eqnarray}
H_1=\sum_{ij} J^{\times}_{ij} (S^x_iS^y_j-S^x_jS^y_i)
\end{eqnarray}
Here the nearest and next-nearest
neighbor cross coupling constant 
are anti-symmetric $J^{\times}_{ij}=-J^{\times}_{ji}$
and are zero for directions perpendicular to the 
motion of the environment $\mathbf{v_n}$.
The term yields an additional
matrix $M_1$ in he random-phase approximation:
\begin{eqnarray}
M_1=(J^{\times}_1(k)+J^{\times}_2(k))\left(
\begin{array}{rrr}
\langle S^z\rangle &0&0 \\
0&\langle S^z\rangle&0 \\
-\langle S^x \rangle&0&0 
\end{array}
\right)
\end{eqnarray}
The matrix is reduced to dimension $3\times 3$ because  we are only
interested
in the superfluid phase only where no difference between the two
sub-lattices is made. The cross coupling terms $J^{\times}$ are given by:
\begin{eqnarray}\label{mgf}
J^{\times}_1=\sum_{a_{AB}} J^{\times}_{AB} \exp(i k a_{AB})\nonumber\\
J^{\times}_1=\sum_{a_{AA}} J^{\times}_{AA} \exp(i k a_{AA}),
\end{eqnarray}
 where $a_{AA}$ and $a_{AB}$ are the lattice parameters
corresponding to A and B lattice sites.
The dispersion relations, given by the eigenvalues
of the total matrix $M+M_1$, are altered in the following way:
\begin{eqnarray}
\omega_k \rightarrow (J^{\times}_1(k)+J^{\times}_2(k))+\omega_k
\end{eqnarray}
in the $k\rightarrow 0$ limit this accounts for a 
tilt of the dispersion curve toward the motion of the 
environment;
the quasi-particle energy in the direction of the motion $v_n$
is lowered while the energy for particles traveling in opposite
directions is lifted.
From the definition of the lattice generating functions $\gamma_1(k)$
and $\gamma_2(k)$
and Eq. (\ref{mgf}) we see that 
$J^{\times}_1(k)=J^{\times}_2(k)=0$
for k where  $\gamma_1(k)=-1$ and $\gamma_2(k)=1$. Hence
the roton dip that triggers the superfluid to supersolid
transition is not affected by the superflow.\newline
The situation is likewise  for the roton minimum at
[111] where $\gamma_1(k)=0$ and $\gamma_2(k)=-1$. 
Also, here the cross couplings $J^{\times}_1$ and  $J^{\times}_2$
become zero and the stability is not affected by induced superflow.
\section{Quantum Fluctuations at Zero Temperature}\label{sec:qfluc}
\begin{figure}
\centering
\resizebox{8.5 cm}{!}{
\includegraphics*{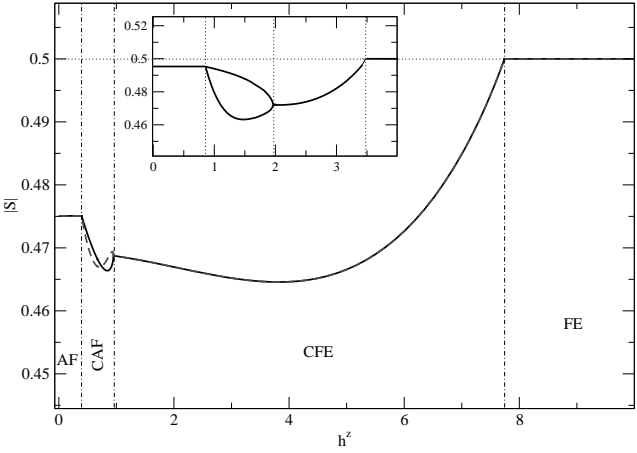} }
\caption{\label{fig:five} The total spin magnitude at zero temperature
is shown for all four phases, for the cases $J^{\top}_1=1.498 K$,
$J^{\top}_2=0.562K$, $J^{\|}_1=-3.899K$ and $J^{\|}_2=-1.782K$, main
plot and $J^{\top}_1=0.5 K$, $J^{\top}_2=0.5 K$, $J^{\|}_1=-2 K$ and 
$J^{\|}_2=-0.5  K$, inset, respectively. The mean-field solution
gives always the horizontal dotted line. In RPA the total spin
is depleted due to quantum fluctuations. For the supersolid (CAF)
phase the RPA gives a difference between the sublattices A and B
(solid and dashed lines main plot). For details, see text.}
\end{figure}

As presented in Ref. \cite{EPL} quantum fluctuations are important
to study as they can lead to the understanding of the physical origin
of the different phases. 
At zero temperature there are no thermal fluctuations
present in the system and all fluctuations will stem from
quantum mechanical effects. The mean-field approximation 
as derived in the beginning of the paper is 
a classical approximation and as such it does not display
quantum fluctuations. This is expressed by 
a constant
spin magnitude of 1/2 over all phases at zero temperature.
In the anisotropic Heisenberg model quantum fluctuations
are a result of non-vanishing pair correlations 
$\langle S^{\mu}_i S^{\nu}_j \rangle-
 \langle S^{\mu}_i  \rangle \langle S^{\nu}_j \rangle $
of nearest and next nearest neighbors. Consequently,
as random-phase approximation takes those correlations accurately into 
account we expect quantum fluctuations which are 
expressed in a depletion of the total spin magnitude as
can be seen in Figure \ref{fig:five}. In the ferromagnetic phase
the total spin is 1/2 and thus there are no quantum fluctuations
present. This is expected as the ferromagnetic phase
is governed in the $h^z\rightarrow \infty$ limit by 
an effective single operator, and hence the classical Hamiltonian:
\begin{eqnarray}
H=\sum_{i} h^z S^z_i
\end{eqnarray} 
The spin depletion is strongly pronounced in the 
CFE and CAF phases where transversal components
account for additional fluctuations.
We also see that in the CAF phase the spin magnitude
is different on the two sub-lattices. This 
indicates that next nearest neighbor interactions are
dominant and balancing nearest neighbor integrations
are slightly suppressed. Therefore we assume that in 
the canted anti-ferromagnetic phase the two sub-lattices do slightly decouple.
This interpretation is supported by the spin-wave excitation
spectrum in the canted anti-ferromagnetic phase. We have seen that there exists
a zero frequency mode, where the spins on different
sub-lattices are $\pi$-phase shifted. 
Hence, spin-waves which couple both  
sub-lattices carry no energy.
\section{Discussion}\label{sec:disc}
\subsection{Case 1}
\begin{figure}
\centering
\resizebox{8.5 cm}{!}{
\includegraphics*{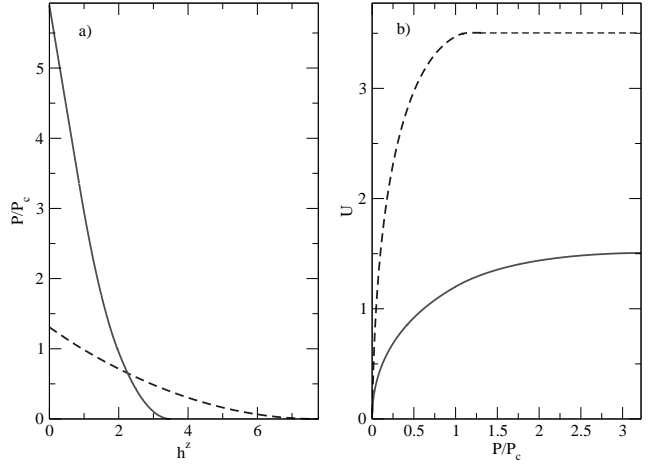} }
\caption{\label{fig:mup} Figure a)  shows the relation between 
pressure in the QLG model and the external magnetic field in the 
anisotropic Heisenberg model. The pressure is normalized with respect to the critical 
pressure $P_c$ marking the supersolid to superfluid transition. 
The dashed line refers to set 1: $J^{\top}_1=1.498 K$,
$J^{\top}_2=0.562 K$,
$J^{\|}_1=-3.899 K$ and
$J^{\|}_2=-1.782 K$  and the solid line to
set 2: 
$J^{\top}_1=0.5 K$,
$J^{\top}_2=0.5 K$,
$J^{\|}_1=-2.0 K$ and
$J^{\|}_2=-0.5 K$ 
Figure b) depicts the internal energy of the model as a function of the pressure
for set 1 (dashed line) and set 2 (solid line). }
\end{figure}
\begin{figure}
\centering
\resizebox{8.5 cm}{!}{
\includegraphics*{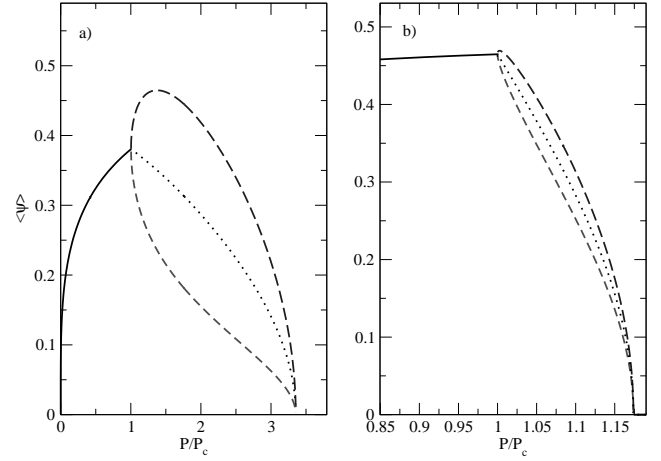} }
\caption{\label{fig:sfofop} Magnitude of the superfluid order parameter 
$\langle \psi \rangle$ in the supersolid and superfluid phase. Long dashed line refers
to the on-site sub-lattice A while the dashed line refers to the interstitial sub-lattice B.
The dotted line is the average of both. Figure a) refers to set 1 and Figure b) to set 2 (see text).}
\end{figure}
\begin{figure}
\centering
\resizebox{8.5 cm}{!}{
\includegraphics*{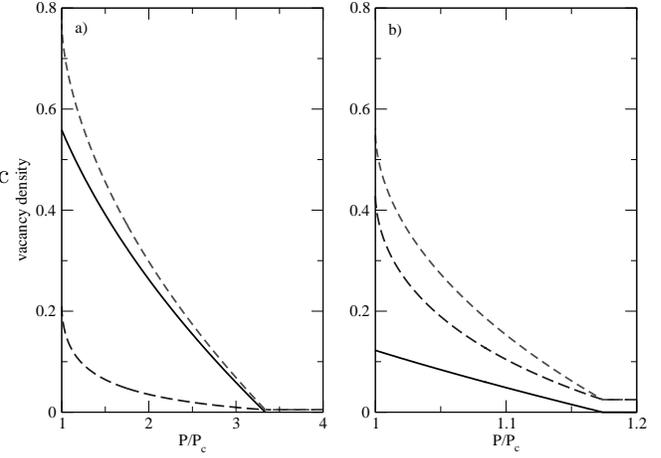} }
\caption{\label{fig:nvdfop} 
Density of vacancies (dashed line), density of interstitials (long dashed line) 
and the difference of both, the net vacancy density (solid line)  in the normal solid 
(for $P/P_c>1$) and the supersolid phases (below 1).
Figure a) refers to set 1 and Figure b) to set 2 (see text).}
\end{figure}

In this section we will discuss the properties of the supersolid phase
using the example of two sets of coupling constants. 
As we are interested in describing real systems we may ask 
what sets of parameters are physical and which set exhibits
the best fit to ${}^4$He.
Physically, the transversal
constants $J^{\top}$ ought to be positive as they correspond to the 
kinetic energy.  
In quantum lattice gas models usually $J^{\top}$s are chosen so that 
the kinetic energy is isotropic up to 4th order giving the best
possible approximation to the continuum limit. 
However in the supersolid phase the Hamiltonian may be regarded
as an effective model and therefore we refrain from this restriction.
The interactions between the helium atoms are controlled
by van-der-Waals forces and their repulsive nature at very short distances
determines negative nearest neighbor interaction $J^{\|}_1$, evoking
anti-ferromagnetic ordering in the spin language. The corresponding 
Lennard-Jones potential is short ranged and therefore it is sufficient 
to only consider nearest and next nearest neighbor interactions.
Liu and Fisher chose coupling constants in order to fit the model
to the actual phase diagram of Helium-4. As the supersolid phase 
had not been discovered experimentally at this time they investigated the 
possibility of a supersolid phase existing. \newline
It is widely accepted that the lambda transitions falls into the universality
class of the XY-model, which is a limiting case of the anisotropic Heisenberg model.
In the same way we believe that the anisotropic Heisenberg model is capable of covering 
the essential properties of the supersolid phase.  
Nevertheless the present model will fail to appropriately 
map ${}^4$He over the complete
range of temperature and pressure as various properties  
such as variability of the lattice constant
and lattice vibration modes (phonons) are
not taken into account in this model. 
Therefore we abstain from fitting the solutions of 
random-phase approximation to the 
phase diagram of real ${}^4$He but merely
choose two sample sets  to
study key properties of the supersolid phase.
\newline
The first set of parameters is given by:
\begin{eqnarray}
J^{\top}_1=0.5 K\nonumber \\
J^{\top}_2=0.5 K\nonumber \\
J^{\|}_1=-2 K\nonumber \\
J^{\|}_2=-.5  K
\end{eqnarray}
In the classical mean-field this set of parameters exhibits all four 
phases where the corresponding critical magnetic fields are given by Eq. 
(\ref{tpfsf}), Eq. (\ref{tpisf}) and Eq. (\ref{tpiaf}) and yield respectively: $h^z_{FE-CFE}=3.5$, $h^z_{CFE-CAF}=2.0207$ 
and $h^z_{CAF-AF}=0.86608$.
In the classical mean-field as well as in the random-phase approximation
the transitions are determined by Equation
(\ref{mf2}), which shows that $h^z$ roughly scales 
with $S^z_A$ and $S^z_B$. Therefore
we expect that the transitions in the random-phase approximation due
to depletion of the 
spin magnitude are slightly lower. The actual values are:
$h^z_{FE-CFE}=3.5$, $h^z_{CFE-CAF}=1.96$ 
and $h^z_{CAF-AF}=0.857$.\newline
The second set we have chosen was extensively studied by  Liu and Fisher and 
their parameters are given by:
\begin{eqnarray}
J^{\top}_1=1.498 K \nonumber \\
J^{\top}_2=0.562 K\nonumber \\
J^{\|}_1=-3.899  K\nonumber \\
J^{\|}_2=-1.782  K 
\end{eqnarray}
Again the transition points are slightly lower (except for 
the FE-CFE transition) for the random-phase approximation and yield:
$h^z_{F-CFE}=7.741 (7.741)$, $h^z_{CFE-CAF}=1.0071 (1.0577)$ 
and $h^z_{CAF-AF}=0.3963 (0.41716)$,
where the numbers in parentheses are the values derived by the classical 
mean-field approximation.
Figure \ref{fig:mup} a) depicts the relation between the external field
of the anisotropic Heisenberg model and the pressure in the QLG 
as given by Equation (\ref{pressa}). The pressure on the y-axis is 
renormalized such that 1 corresponds to the critical pressure $P_c$ of
the superfluid to supersolid transition, given by roughly 20 atm in  Helium 4.
High magnetic field corresponds to low pressure and vice versa.
Negative magnetic fields correspond to high pressures that 
do not have physical validity in the quantum lattice gas. Therefore
the maximal pressure  corresponds to zero magnetic fields.
\newline
In Figure  \ref{fig:mup} b) we plotted the internal energy of
the anisotropic Heisenberg model which corresponds to the 
conjugated potential  $U[\mu]=U-\mu N$ per volume of the QLG model
which is minimized at zero temperature. 
In Figure \ref{fig:sfofop} we plotted the superfluid order
 parameter $\langle \psi \rangle$
as a function of the pressure in the superfluid
 and supersolid phase for both sets of parameters.
The order parameter displays its maximum value
in the vicinity of the transition to the superfluid phase and evidently
approaches zero at the NS-SS transition.
The superfluid order parameter on the on-site 
sub-lattice associated with vacancies is higher
than the one on the interstitial sub-lattice.
While this effect is strongly pronounced in Set 1,
where the order parameter of the vacancies is 
around 37 times the order parameter of the interstitials near
the SS-NS transition, in Set 2 the Bose condensation of the
vacancies is only marginally higher (1.3 times)  than of the 
interstitials. Yet we observe that in this model Bose condensation
appears in the vacancies as well as in the interstitials though 
the major contribution comes from the vacancies.
In Figure \ref{fig:nvdfop} we have the density of vacancies, the 
interstitials and the difference of both, the net vacancy density
plotted as a function of the pressure in the supersolid 
and normal solid phases.
We see that in the normal solid phase the number of vacancies 
and interstitials stays finite. This is due to quantum fluctuations 
and consequently the number of vacancies have in addition to thermal activation
a second contribution resulting from quantum mechanical effects.
Nevertheless the number of vacancies and interstitials appear in equal 
numbers and the net contribution in the normal solid is zero. 
This is different in the supersolid phase where a surplus of vacancies
accounts for  a positive net vacancy density.
As we decrease the pressure in the supersolid phase both 
the vacancy density and the 
interstitial density increase. However, the vacancy density 
increases faster, leaving 
a net vacancy density which reaches its maximum at the 
phases transition between the 
supersolid and the superfluid.
Interestingly the net vacancy density varies nearly linearly with the pressure
as the solid line in Figure  \ref{fig:nvdfop} shows.
\subsection{Case 2}
In the section on the excitation spectrum we have seen that 
the spin-wave energy at [100] of the first Brillouin zone goes soft
exactly when the superfluity to supersolid phase transition  occurs.
Additionally, for coupling constants that fulfill condition Eq. (\ref{iscr2})
there exists a second minimum at [111]  which can collapse.
Following set of constants fulfill this condition:
\begin{eqnarray}
J^{\top}_1=0.5K \nonumber \\
J^{\top}_2=0.5K \nonumber \\
J^{\|}_1=-2K \nonumber \\
J^{\|}_2=-1.5K  
\end{eqnarray}
According to Equation (\ref{tpfsf}) there is one (normal fluid to superfluid)
transition in the system:
\begin{equation}
h^z_{FE-CFE}=4.5
\end{equation}
beneath this line classical mean-field approximation predicts a  CFE  (superfluid)
phase that extends down to $h^z=0$ as 
due to the relatively large negative $J^{\|}_2$
the solid phase does not acquire a sufficiently low 
free energy to be the true ground state. 
The random-phase approximation however draws a slightly different picture.
Analogous to the classical mean-field solution the random-phase approximation also
yields a phase transition near $h^z=4.5$.
\begin{figure}
\centering
\resizebox{8.5 cm}{!}{
\includegraphics*{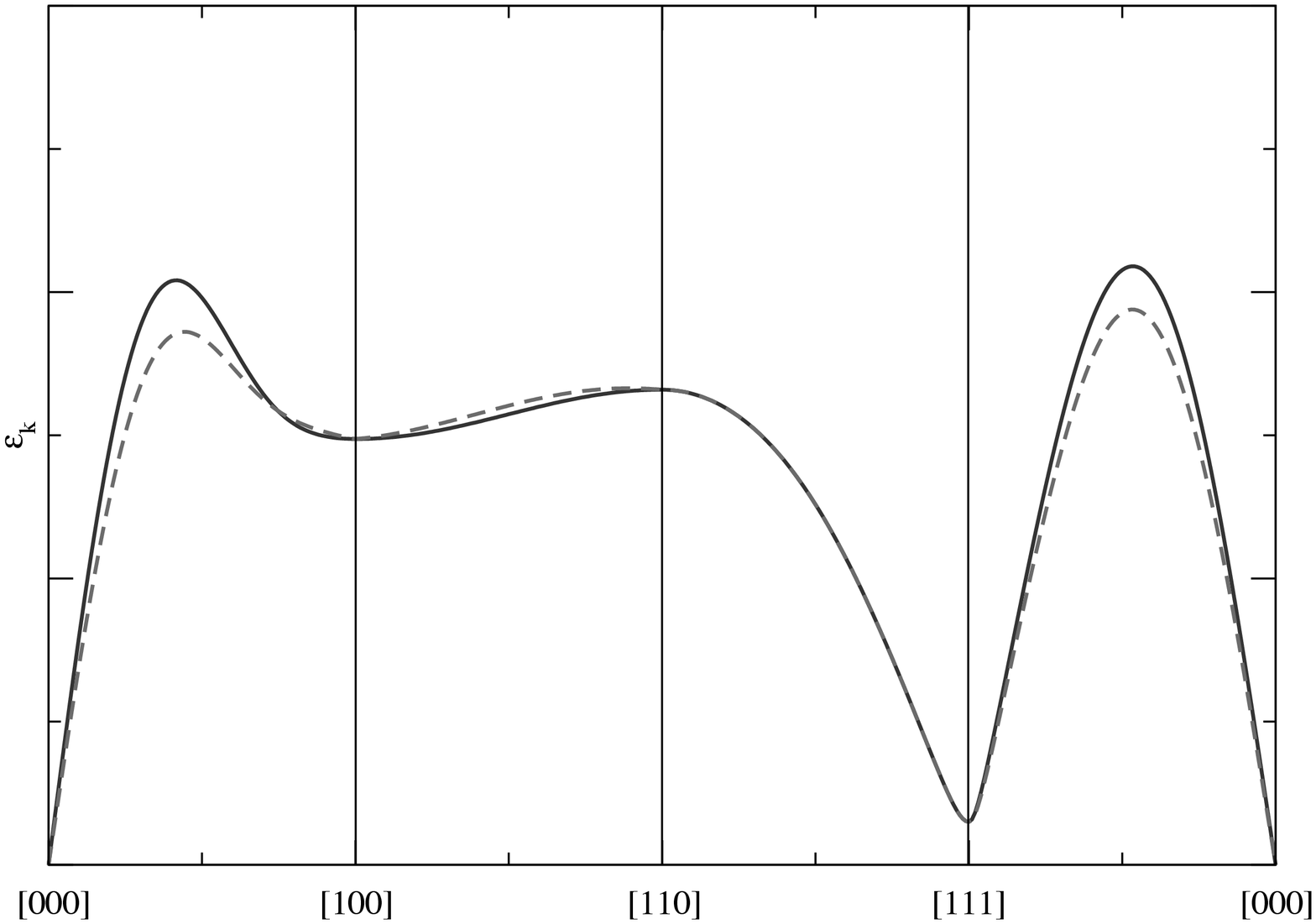} }
\caption{\label{fig:c3pd} Excitation spectrum (solid line) in the 
superfluid phase for $J^{\top}_1=0.5K$,$J^{\top}_2=0.5K$,
$J^{\|}_1=-2K$  
$J^{\|}_2=-1.5K$ just before the phase become unstable due
to a collapsing minimum at [111]. Induced superflow 
alters the spectrum (dashed line) but does not affect the minimum.  }
\end{figure}
But unlike the classical mean-field solution, the superfluid phase here does not
survive all the way down to $h^z=0$. 
Due to the particular choice of parameters 
the superfluid phase becomes 
unstable at around $h^z=2$; i.e the quasi-particle spectrum 
turns imaginary at
$\gamma_1(k)=0$ and $\gamma_2(k)=-1$ ([111]), as
Figure \ref{fig:c3pd} shows. The dashed line in this figure shows 
the excitation spectrum under an induced superflow $v_s$. The
roton minimum is not affected by this superflow and hence
the superfluid phase is not destabilized by an spontaneously induced superflow. 
Interestingly beyond this line no other 
stable phase exists in the random-phase approximation;
there is no set of spin fields $\langle S^x_A\rangle$,
 $\langle S^x_B\rangle$,
  $\langle S^z_A\rangle$ 
and $\langle S^z_B\rangle$ that 
solves the self-consistency equations of the random-phase approximation. 
Therefore we conclude that there must exist a 
'novel' phase that is not covered by the random-phase approximation on a
bipartite lattice and we will leave the detailed discussion of this phase to future work. 
\section{Conclusion}
In this paper we analysed the supersolid phase in the three 
dimensional quantum lattice gas model. Through transformation to the 
anisotropic Heisenberg model in a external field we were 
able to employ the well-established technique of real-time Green's functions
for spin systems. The series of infinite order Green's functions as it appears 
in the equation of motion was truncated by applying cumulant decoupling and
the resulting random-phase approximation accounts for linear spin-waves. 
We are the first to apply this method
to the canted anti-ferromagnetic phase entailing a set of 6 algebraic 
equations.  The innate self-consistency equations inhere a 3 dimensional 
numerical integral over the k-space. By introducing a two dimensional 
density of states  the integral was reduced to 
two dimensions where the lattice generating functions serve as new
integration variables. In the said integral the DOS is the only
quantity that depends on the structure of the lattice. Hence, once
the DOS is computed for a certain lattice geometry the further
calculation remain unaltered. Therefore our method is widely applicable
and easily adjustable to various magnetic systems where 
canted phases are in the center of interest. This also holds for
2 dimensional lattices where linear spin waves are expected to yield a
reasonable approximation. \newline
The random-phase approximation takes quantum fluctuations into account and 
consequently in this solution the solid phase exhibits vacancies and
interstitials at zero temperature.
Yet in the normal solid phase the vacancies and the interstitials occur in
equal number, thus yielding a zero net vacancy density.
In the supersolid phase this balance shifts in favor of the vacancies giving rise
to a finite positive net vacancy density at zero temperature.  Our data also shows that 
vacancies as well as interstitials Bose condense and hence both contribute to 
superfluidity.  Nevertheless the Bose condensation is stronger expressed  in the vacancies
thus giving the major contribution to supersolidity. \newline 
Furthermore the present approach confirms suggestions that the superfluid 
to supersolid transition is triggered by a collapsing roton minimum. However 
our results show that this roton dip is not affected by 
Galilean transformation and hence the superfluid phase is stable against spontaneously induced
superflow. Additionally we find that for a narrow regime of parameters a second 
roton minimum collapses. Below this point a stable phase does not exist
in the bipartite random-phase approximation and a solution is thus beyond the model.
The prospect of future work looks promising. The formalism is easily
extendable to finite temperatures as shown in Ref. \cite{finite}, where
we investigated the properties of the supersolid phase at finite T. 
In particular the temperature dependence of the net vacancy density and the 
behavior of the specific heat across the supersolid to normal solid
transitions is of particular interest.


\begin{thebibliography}{99}          

\bibitem{andreev} A. F. Andreev and I. M. Lifshitz, Zh. Eksp. Teor. Fiz. 56, 2057 (1969)  [JETP 29, 1107 (1969)].

\bibitem{chester} G. V. Chester, Phys. Rev. A 2, 256 (1970). 

\bibitem{leggett} A. J. Leggett, Phys. Rev. Lett. 25, 1543 (1970).  

\bibitem{kim1} E. Kim, M.H.W. Chan, Nature 427, 225 (2004).

\bibitem{kim2} E. Kim, M.H,W. Chan, Science 305, 1941 (2004).

\bibitem{aoki} Y. Aoki, J. C. Graves, H. Kojima, Phys. Rev. Lett. 99, 015301 (2007).  

\bibitem{day} Day, J.R. Beamish, J. Nature 450, 853856 (2007). 

\bibitem{anderson} P.W. Anderson, Nature Phys. 3, 160 (2007). 

\bibitem{superglass} M. Boninsegni, N. Prokof'ev, B. Svistunov, Phys. Rev. Lett. 96, 105310 (2006). 

\bibitem{fisher1} K.-S. Liu, M.E. Fisher, J. Low. Temp. Phys. 10, 655 (1973). 

\bibitem{rotonmin} P. Nozi\`eres, J. Low Temp. Phys. 137, 45 (2004).

\bibitem{zhao} E. Zhao, A. Paramekanti, Phys. Rev. Lett. 96, 105303 (2006).

\bibitem{EPL} A. Stoffel and M. Gulacsi, Europhys. Lett. {\bf 85}, 20009 (2009). 

\bibitem{matsubara} T. Matsubara and H. Matsuda,Progr. Theoret. Phys. 16, 569 (1956);17, 19 (1957). 

\bibitem{matsuda} H. Matsuda, T. Tsuneto, Prog. Theoret. Phys. Suppl. 46, 411 (1970).

\bibitem{fisher}M.E. Fisher, Rep. Prog. Phys. {\bf 30}, 615 (1967). 

\bibitem{bogo} N.N. Bogolyubov, S.V. Tyablikov, Doklady Akad. Nauk. S.S.S.R. 126, 53 (1959)
 [translation: Soviet Phys. -Doklady 4, 604 (1959)].

\bibitem{tyablikov} S.V. Tyablikov, Ukrain. Mat. Yhur. 11, 287 (1959).

\bibitem{auer} A. Auerbach, {\sl{Interacting electrons and quantum magnetism}}, Springer, (1994).

\bibitem{brown}P.E. Bloomfield and E.B. Brown, Phys. Rev. B{\bf{22}}, 1353 (1980). 

\bibitem{nafari}P.E. Bloomfield and N.Nafari, Phys. Rev. A{\bf{5}}, 806 (1972).

\bibitem{finite} A. Stoffel and M. Gulacsi, Euro. Phys. J. B{\bf 67}, 169 (2009).

\end{thebibliography}
\end{document}